\documentclass[pdflatex,prl,twocolumn,showpacs,superscriptaddress,nofootinbib]{revtex4-2}
\usepackage{bm}
\usepackage{color}
\usepackage{xcolor}
\usepackage{soul}
\usepackage{hyperref}
\usepackage{physics}
\usepackage{placeins}
\usepackage{amsmath,amssymb}
\usepackage{graphicx,multirow,tabularx}
\usepackage{davitex}
\usepackage{comment}
\usepackage{multirow}
\usepackage{siunitx}
\usepackage[ruled,vlined]{algorithm2e}
\usepackage{lineno}
\newcommand \pade {Pad\'e~}

\begin{document}
\title{Searching for the QCD critical endpoint using multi-point Pad\'e approximations}

\author{D. A. Clarke}
\affiliation{Department of Physics and Astronomy, University of Utah,
Salt Lake City, Utah, United States}
\author{P. Dimopoulos}
\affiliation{Dipartimento di Scienze Matematiche, Fisiche e Informatiche, Università di Parma and INFN, Gruppo Collegato di Parma
I-43100 Parma, Italy}
\author{F. Di Renzo}
\affiliation{Dipartimento di Scienze Matematiche, Fisiche e Informatiche, Università di Parma and INFN, Gruppo Collegato di Parma
I-43100 Parma, Italy}
\author{J. Goswami}
\affiliation{RIKEN Center for Computational Science, Kobe 650-0047, Japan}
\author{C. Schmidt}
\affiliation{Universität Bielefeld, Fakultät für Physik, D-33615 Bielefeld, Germany}
\author{S. Singh}
\affiliation{Universität Bielefeld, Fakultät für Physik, D-33615 Bielefeld, Germany}
\author{K. Zambello}
\affiliation{Dipartimento di Fisica dell’Università di Pisa and INFN--Sezione di Pisa, 
Largo Pontecorvo 3, I-56127 Pisa, Italy.}
\date{\today}

\newcommand{\muCEP}{\mu_B^{\text{CEP}}}
\newcommand{\TCEP}{T^{\text{CEP}}}
\newcommand{\tLYE}{t^{\text{LYE}}}
\newcommand{\hLYE}{h^{\text{LYE}}}

\begin{abstract}

Using the multi-point Pad\'e approach, we locate Lee-Yang edge singularities of the QCD pressure
in the complex baryon chemical potential plane. These singularities are extracted from singularities
in the net baryon-number density calculated in $N_f=2+1$ lattice QCD at physical quark mass
and purely imaginary chemical potential. Taking an appropriate scaling ansatz in the vicinity
of the conjectured QCD critical endpoint, we extrapolate the singularities on
$N_\tau=6$ lattices to pure real baryon chemical potential to estimate
the position of the critical endpoint (CEP). We find $\TCEP=105^{+8}_{-18}$~ MeV and $\muCEP = 422^{+80}_{-35}$~ MeV, which compares well with recent estimates in the literature.
For the slope of the transition line at the critical point we find $-0.16(24)$.
\end{abstract}

\pacs{11.10.Wx, 11.15.Ha, 12.38.Aw, 12.38.Gc, 12.38.Mh, 24.60.Ky, 25.75.Gz, 25.75.Nq}

\maketitle

\textit{Introduction}.---A central goal of the experimental program at the Relativistic Heavy-Ion Collider (RHIC) 
of Brookhaven National Laboratory (BNL) in the US and at the Large Hadron Collider (LHC) at CERN, 
Switzerland is the exploration of the phase diagram of quarks and gluons in the plane
of temperature $T$ and baryon chemical potential $\mu_B$ as described 
by the theory of the strong interaction, quantum chromodynamics (QCD). 
At low $T$ and $\mu_B$, QCD matter
is known to exist as a gas of hadrons. At high $T$ and/or $\mu_B$, hadrons start to melt
and quark-gluon plasma (QGP) is created.  The QGP created at high $\mu_B$ 
may experience a sharp first-order phase transition as it cools, with bubbles of 
QGP and hadrons coexisting at a well-defined temperature. 
The coexistence region ends in a critical point (CEP), where QGP and hadronic 
matter become indistinguishable.
The conjectured CEP belongs to the 3-$d$, $Z(2)$
universality class.

Progress in understanding the phase diagram at $\mu_B>0$ from first-principle lattice QCD 
calculations, in particular locating the CEP, is stymied by the infamous sign problem. 
In spite of this difficulty, lattice calculations are able to provide some controlled 
information of the diagram at sufficiently small $\mu_B/T$. This is accomplished
using various techniques, for instance reweighting~\cite{Barbour_1998,Fodor_2002}, 
analytic continuation from
purely imaginary $\mu_B$~\cite{deForcrand:2002hgr,DElia:2002tig}, 
and Taylor expansion in $\mu_B/T$~\cite{Gavai:2003mf,Allton:2005gk}. 
The Taylor expansion technique, while highly successful, is severely limited 
by the computational power
required to compute higher-order Taylor coefficients, with state-of-the-art
calculations achieving eighth order~\cite{Bollweg:2022rps,Borsanyi:2018grb}.
In response to this challenge, various resummation schemes have been 
proposed~\cite{Borsanyi:2021sxv,Mondal:2021jxk,Mitra:2022vtf},
which all attempt to probe deeper into the phase diagram without having
to compute even higher-order cumulants. 
Recently the STAR collaboration has found tantalizing evidence of the hint of a QCD critical point~\cite{STAR:2020tga} in the net proton fluctuation data. However, from the analysis of the QCD equation of state~\cite{Bollweg:2022fqq,Goswami:2022nuu} using Taylor expansion and \pade approximation, it is concluded that the CEP is likely not located in the energy range of the beam energy scan II (BESII) program at RHIC in collider mode.

\begin{figure}
\centering
\includegraphics[width=0.483\textwidth
]{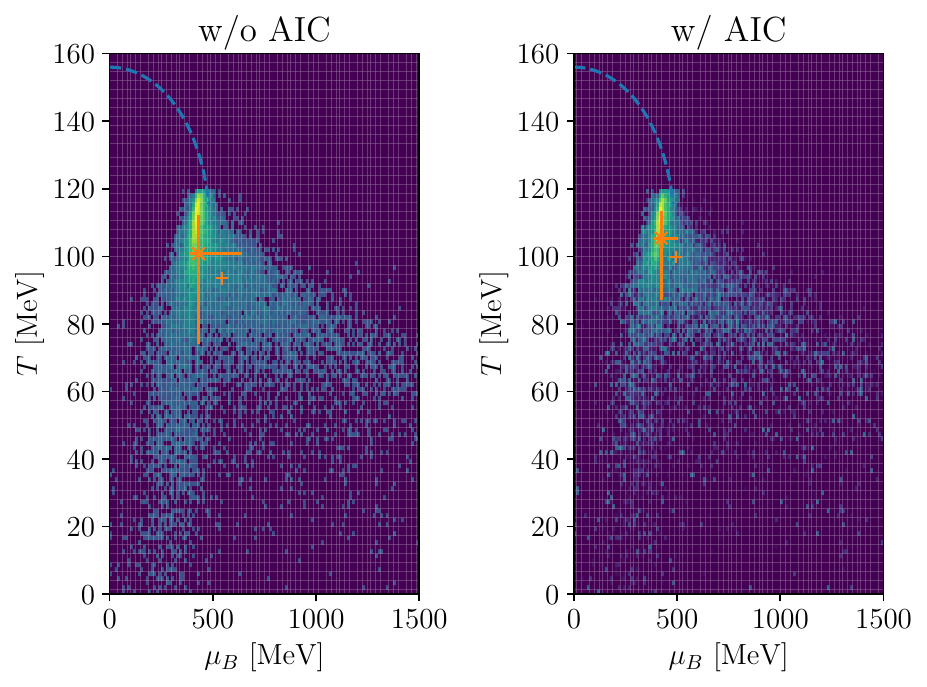}
\caption{Probability distribution of the QCD critical point from extrapolating Lee-Yang singularities to the real domain using universal scaling. For a detailed description see the text.}
\label{fig:CEP_histo}
\end{figure}

In this paper we adopt the multi-point-Pad\'e resummation method introduced in~\cite{Dimopoulos:2021vrk}. The method uses information from simulations at purely imaginary $\mu_B$ to construct a Pad\'e approximation to the logarithm of the QCD grand partition function $\log\ZQCD$ for complex $\mu_B$. 
We then determine singularities of the approximant to estimate the CEP location. 
In particular we consider temperature-like and magnetization-like couplings $t$ and $h$ near the CEP. 
Then according to the Lee-Yang theorem~\cite{Yang:1952be} applied to the universal theory (3-$d$ $Z(2)$), at $t=0$, zeroes of $\ZQCD$ in the complex $h$-plane that approach the real axis in the thermodynamic limit correspond to phase transitions. 
For $t>0$, the closest singularities to the origin are the Lee-Yang edges (LYE). We extrapolate the position of the singularity to the the CEP by following LYE scaling~\cite{Fisher:1978pf}. 

From model calculations and model-independent symmetry arguments we understood that the location of $\TCEP$ has to be searched below the chiral transition temperature $T_c^{\rm chiral}\approx 132$~MeV \cite{HotQCD:2019xnw}. 
Thus, we extend our calculations down to $T=120$ MeV in this paper.
Our final results are summarized in \figref{fig:CEP_histo}. We note that they are consistent with the bounds on $T$ and $\mu_B$ mentioned above. 

\textit{Lattice simulation details.}---We generated configurations for a $N_\sigma^3\times N_\tau$
lattice with $N_\sigma=36$ and $N_\tau=6$ and $N_f=2+1$ dynamical highly improved staggered quarks (HISQ)~\cite{Follana:2006rc} using the \simulat~\cite{Altenkort:2021cvg,HotQCD:2023ghu} implementation of the rational hybrid Monte Carlo algorithm (RHMC) \cite{Clark:2006fx}. 
We choose bare parameters along the line of constant physical pion mass obtained for the ratio of light to strange quark mass $m_s/m_l=27$ in Refs.~\cite{Bazavov:2011nk, HotQCD:2014kol, Bollweg:2021vqf}. 
The simulations run at $\mathcal{O}(10)$ pure imaginary $\mu_B$ in the range $\mu_B/T\in[0,i\pi]$ to avoid the sign problem. 
For simplicity we set light and strange quark chemical potentials to equal values, $\mu_l=\mu_s$, which corresponds to baryon chemical potential $\mu_B=3\mu_l$ and zero strangeness chemical potential $\mu_S=0$. 
A number of configurations ranging from $1800$ to $24000$ was generated for a set of five temperatures ($T = 166.6, 157.5, 145.0, 136.1$ and $120.0$ MeV) extending far below the chiral transition temperature as summarized in \tabref{tab:latticeSetup}. 
The configurations are separated by 10 molecular dynamic time units (MDTU).

\begin{figure}
    \centering
     \includegraphics[width=1.0\linewidth]{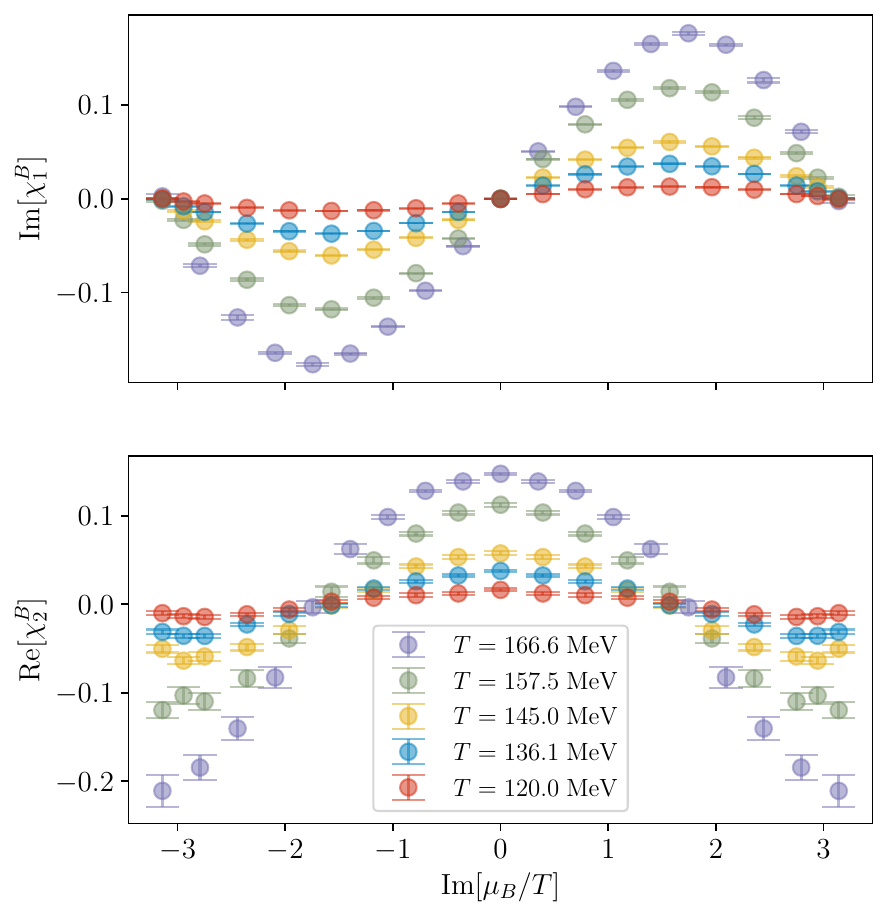}
    \caption{First (top) and second (bottom) order net baryon-number cumulants at $T = 166.6, 157.5, 145.0, 136.1$ and $120.0$ MeV. The points in the interval $[0,i\pi]$ are calculated, the points in $[-i\pi,0]$ are obtained by reflection symmetry.
    }
    \label{fig:chiral_obs}
\end{figure}

We measured the first- and second-order cumulants of the net baryon-number density, defined as
\begin{equation}
    \chi_n^B = \frac{N_\tau^3}{N_\sigma^3} \left( \frac{\partial}{\partial \hat{\mu}_B} \right)^n \log\ZQCD.
\end{equation}
A total of 500 random vectors were used to construct unbiased noisy estimators of the observables.
Numerical results are illustrated in \figref{fig:chiral_obs}. The first-order cumulant is a pure imaginary, odd, and $2\pi$-periodic function of $\mu_B/T$ whose signal gets damped as the temperature is decreased. Conversely the second-order cumulant is a pure real, even, and $2\pi$-periodic function of $\mu_B/T$.

\begin{table}
\begin{tabularx}{\linewidth}{LCCR}
\hline\hline
  $\beta$ & $T$ [MeV] & $N_\mu$ &$\nconf/N_\mu$\\
\hline
  6.170  &    166.6 & 10 & 1800   \\
  6.120  &    157.5 & 10 & 4780    \\
  6.038  &    145.0 & 10 & 5300  \\
  5.980  &    136.1 & 10 & 6840 \\
  5.850  &    120.0 & 10 & 24000   \\
  \hline\hline
\end{tabularx}
\caption{Summary of statistics for each ensemble used in this study. The last column gives
approximate number of thermalized configurations per $\mu$ value.
Quark masses are fixed to their
physical value and $\mu_s=\mu_l$.}
\label{tab:latticeSetup} 
\end{table}

\begin{figure}
    \includegraphics[width=1.0\linewidth]{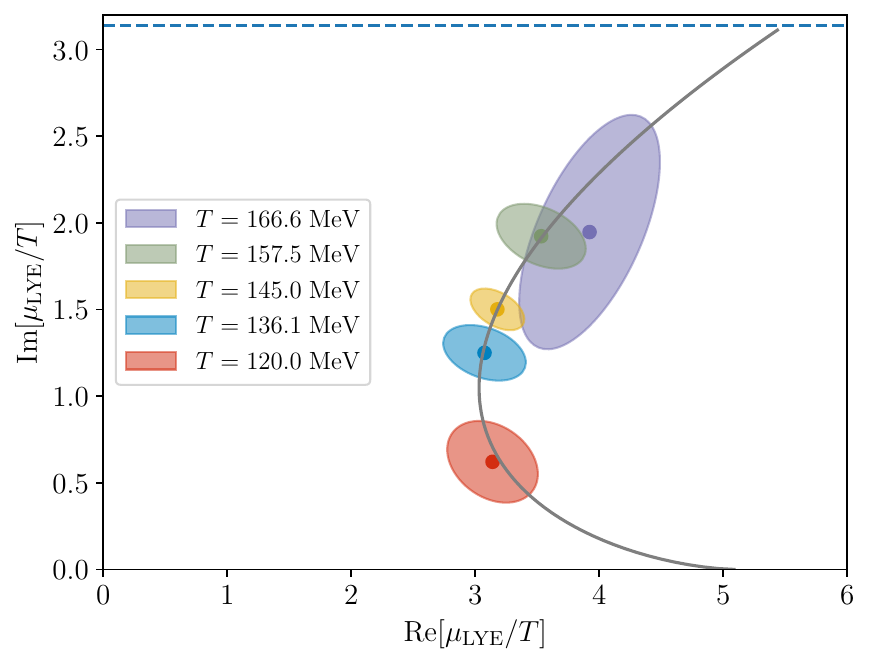}
    \caption{Singularities at $T = 166.6, 157.5, 145.0, 136.1$ and $120.0$ MeV. The dashed line lies at $\hat{\mu}_B = i\pi$.}
    \label{fig:confelipse}
\end{figure}

\textit{Mapping Poles of the multi-point {\pade} to Lee-Yang Edge singularities.}---The first order cumulant is approximated by a rational function of the form 
\begin{equation}
    R^3_3(x)=\frac{\sum\limits_{i=0}^3a_i x^i}{1+\sum\limits_{i=1}^3 b_ix^i}\,.
    \label{eq:ratapprox}
\end{equation}
In the generalized least square process \cite{Dimopoulos:2021vrk}, data of the first two cumulants are taken into account. 
The fit interval $[\hat\mu_{\text{min}},\hat\mu_{\text{max}}]$ is contained in $[-i\pi,+i\pi]$, and the length $\hat\mu_{\text{max}}-\hat\mu_{\text{min}}$ is varied between $\pi$ and $2\pi$. 
In this way we construct 55 rational approximations per temperature. 
We obtain the poles of the rational approximation by calculating the roots of the polynomial in the denominator. 
We keep only the roots in the first quadrant and pick the one which is closest to the center of our fit interval. 
For each fit interval and temperature we repeat the calculation of poles on $\mathcal{O}(200)$ samples, generated by bootstrapping over the standard deviation of the cumulant data.
In this way we generate distributions of poles which we may represent as 1$\sigma$-confidence ellipses, as shown in \figref{fig:confelipse} for a particular choice of intervals. 
We observe that the imaginary part of the poles decreases with decreasing temperature. 
The large error on the position of the $T=166.6$ point is likely due to the much lower number of gauge configurations as compared to the other points (see \tabref{tab:latticeSetup}). The solid grey line in \figref{fig:confelipse} stems from the fit described below.  
We note that for $T>170$ MeV, the poles accumulate on the dashed line in \figref{fig:confelipse} and follow a scaling associated with the Roberge-Weiss transition in QCD \cite{Dimopoulos:2021vrk,Clarke:2023noy,Schmidt:2024edp}.

\textit{Estimation of CEP location.}---The QCD CEP is expected to belong to the 3-$d$, $Z(2)$ universality class. 
The mapping from the control parameters $T$ and $\mu_B$ to 
the temperature-like and magnetization-like scaling directions $t$ and $h$ of the Ising model is not known. 
We thus adopt a frequently used linear ansatz for the mapping \cite{Rehr:1973zz, Nonaka:2004pg, Parotto:2018pwx, Kahangirwe:2024cny},
\begin{equation}
\begin{aligned}\label{eq:CEPansatz}
t &=A_t\Delta T+B_t \Delta\mu_B,\\
h &=A_h\Delta T+B_h \Delta\mu_B, 
\end{aligned}
\end{equation}
with $\Delta T\equiv T-\TCEP$, $\Delta \mu_B\equiv\mu_B-\muCEP$, and
constants $A_i$, $B_i$.
For the extrapolation of the poles to the real axis, we would like to follow the path of the LYE.  
Expressed in terms of the scaling variable $z\equiv t/|h|^{1/\beta\delta}$, it has a constant position \cite{Fisher:1978pf}
\begin{equation}\label{eq:zLYE}
    \zLYE = |\zLYE|~e^{i\pi/2\beta\delta}.
\end{equation}
Here $\beta$,$\delta$ are the well known critical exponents of the 3-$d$, $Z(2)$ universality class; we use the values $\beta=0.326419$, $\delta=4.78984$ \cite{El-Showk:2014dwa}.
For a discussion of the universal constant $|\zLYE|$ see \cite{Connelly:2020gwa,Rennecke:2022ohx,Johnson:2022cqv,Karsch:2023rfb}.
Plugging \equatref{eq:CEPansatz} into \equatref{eq:zLYE} then implies
that $\muLYE$ should scale~\cite{Stephanov:2006dn} as
\begin{equation}\begin{aligned}\label{eq:CEPfit}
      \Re\muLYE&=\muCEP+ c_1\Delta T +  c_2\Delta T^2 + O(\Delta T^3)  \\
      \Im\muLYE&=c_3\Delta T^{\beta\delta},
\end{aligned}\end{equation}
where the coefficients $c_i$ are functions of the mixing parameters $A_j$, $B_j$. 
In particular we have $c_1=A_h/B_h$, which denotes the slope of the transition line at the CEP in the ($T$,$\mu_B$)-diagram. 
Note that the presence of the coefficient $c_2$ goes beyond the  linear ansatz of \equatref{eq:CEPansatz} but seems important to extrapolate our current data. 

We use \equatref{eq:CEPfit} to simultaneously fit the real and imaginary parts of our singularities. In total the fit has 5 parameters, $\muCEP$, $\TCEP$, $c_1$, $c_2$, and $c_3$. We checked that separate fits to real and imaginary parts give very similar results with slightly reduced errors. 

\begin{figure}
    \centering
    \includegraphics[width=\linewidth]{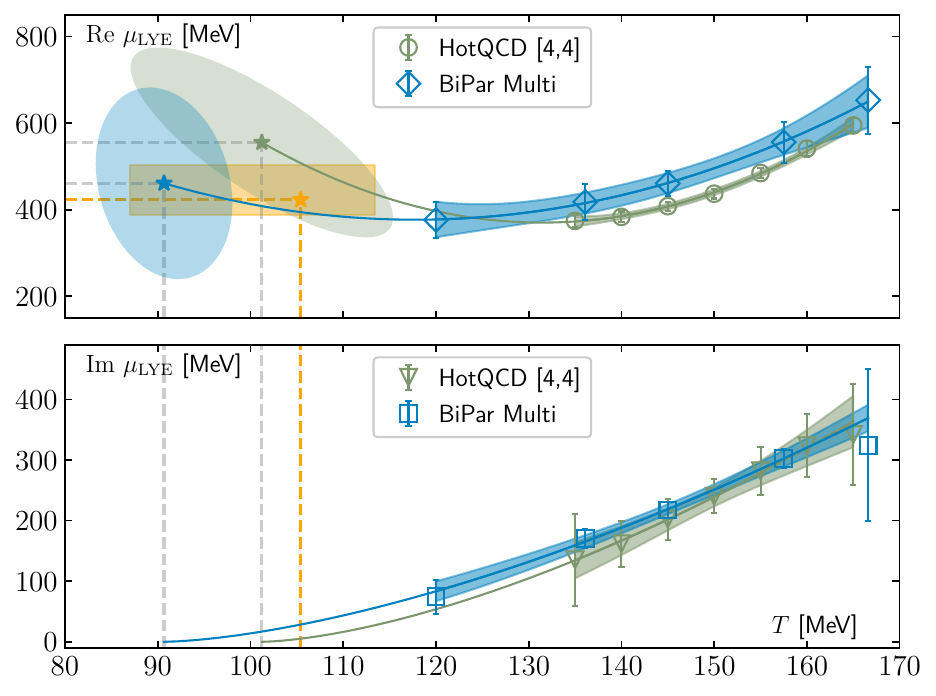}
    \caption{Scaling fits for the LYE singularities related to the CEP. Green data come from a [4,4] Pad\'e from Ref.~\cite{Bollweg:2022rps}.
    Blue data come from the multi-point Pad\'e. 
    {\it Top}: Scaling of the real part. {\it Bottom}: Scaling of the
    imaginary part. The ellipses shown in the top panel represent the 68\% confidence region deduced from the covariance matrix of the fit. The orange box indicates the AIC weighted estimate \eqref{eq:AICweightedEstimate}.}
    \label{fig:CEPextra}
\end{figure}

\textit{Results for the CEP.}---We perform $\mathcal{O}(10^5)$ different fits by varying the rational approximation for each temperature, based on different intervals and bootstrapping over the data. 
The results for the coordinates of the CEP are presented in \figref{fig:CEP_histo} as a histogram, weighted with (right) and without (left) the Akaike information criterion (AIC). The median and the 68\% confidence interval is estimated and presented in \figref{fig:CEP_histo} by star symbols and error bars. 
In the AIC-weighted case we find
\begin{equation}\label{eq:AICweightedEstimate}
    (\TCEP,\muCEP)=(105^{+8}_{-18}, 422^{+80}_{-35})\;\text{MeV}\,,
\end{equation}
where the errors are statistical errors only. The plus symbols represent the arithmetic mean. 
Also plotted in \figref{fig:CEP_histo} as a blue, dashed line is the crossover temperature. 
It is interesting to note that the histograms indicate two branches in the tails of the distribution.
Whether these branches contain further information on the QCD phase diagram, e.g. on binodal or spinodal lines, will be interesting to discuss in future publications. 

In order to compare the $N_\tau=6$ results from imaginary chemical potential calculations with the eighth-order Taylor expansion results from $N_\tau=8$ \cite{Bollweg:2022rps}, the fit with the best (smallest) $\chidof$ is presented in \figref{fig:CEPextra} (and \figref{fig:confelipse}). 
The same fit is applied to the singularities extracted from the [4,4]-\pade resummation of the pressure expansion presented in \cite{Bollweg:2022rps}. 
The blue and green ellipses drawn in \figref{fig:CEPextra} are the 68\%-confidence ellipses obtained from the covariance matrix of the fit parameters. 
The orange square gives the estimate of the AIC-weighted median over all $N_\tau=6$ fits.
The error bands on the fit are obtained with error propagation implemented in the AnalysisToolbox~\cite{toolbox}, which is also used to carry out the fits.
Data points and fits of $N_\tau=6$ and $N_\tau=8$ calculations are compatible within errors; the extrapolated location of the CEP might, however, give rise to some cutoff effects. 
The fit parameters are summarized in Table~\ref{tab:fitParameter}.

\begin{table*}[htbp]
    \centering
    \begin{tabular}{%
    p{1.5cm}
    rcrcr
    rcrcr
    rcrcr
    rcl
    rcl
    rcl
    }\hline\hline
         &  \multicolumn{15}{c}{$\mathbf{N_\tau=6}$} 
         &  \multicolumn{9}{c}{$\mathbf{N_\tau=8}$}\\
         &  \multicolumn{15}{c}{\textbf{multi-point Pad\'e}} 
         &  \multicolumn{9}{c}{\textbf{[4,4]-Pad\'e}} \\ 
         \\[-0.75em]
         & \multicolumn{5}{c}{$\TCEP$ [MeV]}
         & \multicolumn{5}{c}{$\muCEP$ [MeV]}
         & \multicolumn{5}{c}{$\mu_B/T$}
         & \multicolumn{3}{c}{$\TCEP$ [MeV]}
         & \multicolumn{3}{c}{$\muCEP$ [MeV]}
         & \multicolumn{3}{c}{$\mu_B/T$} \\
         \\[-0.75em] \hline
         \\[-0.75em]
         best fit
         & {90.7} & $\pm$ & {7.7} & &
         & $\;\,${461.2}& $\pm$ & {220}\phantom{.0} & &
         & $\;\,${5.09} & $\pm$ & {0.68}& &
         & $\;\,${101}  & $\pm$ & {15}  
         & {560}  & $\pm$ & {140} 
         & {5.5} & $\pm$ & {1.7}       \\
         weight-1
         & $\;\,${105.4} & $+$ & {8.0}  & $-$ & {18.4}
         & $\;\,${422.9} & $+$ & {80.5} & $-$ & {34.9}
         & $\;\,${3.92}  & $+$ & {1.52} & $-$ & {0.24}
         &  &  &
         &  &  &
         &  &  & \\ 
         weight-2
         & $\;\,${100.8} & $+$ & {11.6}  & $-$ & {26.8} 
         & $\;\,${430.9} & $+$ & {208.2} & $-$ & {42.2}
         & $\;\,${4.20}  & $+$ & {4.13}  & $-$ & {0.47}
         &  &  &
         &  &  &
         &  &  & \\ \hline 
                 \\[-0.75em]
         & \multicolumn{5}{c}{$c_1$ }
         & \multicolumn{5}{c}{$c_2$ }
         & \multicolumn{5}{c}{$c_3$ }
         & \multicolumn{3}{c}{$c_1$ }
         & \multicolumn{3}{c}{$c_2$ }
         & \multicolumn{3}{c}{$c_3$} \\
         \\[-0.75em] \hline
         \\[-0.75em]
         best fit  
         & {-6.2} & $\pm$ & {9.2} & &
         & $\;\,${0.115}& $\pm$ & {0.090} & &
         & $\;\,${0.424} & $\pm$ & {0.086}& &
         & $\;\,${-12.3}  & $\pm$ & {8.1}  
         & $\;\,${0.203}  & $\pm$ & {0.059} 
         & $\;\,${0.55} & $\pm$ & {0.25}       \\ \hline \hline
    \end{tabular}
    \caption{Obtained fit parameters from the fit with \equatref{eq:CEPfit} to the real and imaginary parts of the singularities of $\log Z$. For $N_\tau=6$, we show the results for the fit with the smallest $\chidof$ (0.067), as well as the median and $1\sigma$-percentiles of all performed fits, weighted with (weight-1) and without (weight-2) the AIC.}
    \label{tab:fitParameter}
\end{table*}

Besides the location of the CEP, we can also estimate the slope of the transition line at the CEP. In the $\mu_B,T$-diagram, the slope is given as $1/c_1=-0.16(24)$. Since the direction of the temperature-like scaling field $t$ is tangential to the transition line at the CEP, we can also estimate the angles between the $t$-axis and the axes of the $\mu_B,T$-diagram. 
\begin{equation}
\measuredangle (t,T)=80.8^\circ \pm 13.4^\circ\quad  \measuredangle (t,\mu_B)=9.2^\circ \pm 13.4^\circ\;.   
\end{equation}
We note that the map in \equatref{eq:CEPansatz} can be decomposed into translation, rotations and scales. 
In this case one of the above angles enters directly the mapping between the QCD parameter and the scaling fields. 

\textit{Discussion.}---An important consistency check for our results on the CEP location is the comparison with the crossover line. A natural parametrization of the pseudo-critical temperature $T_{pc}$ is given as  
\begin{align}
        T_{pc}(\mu_B) = T_{pc,0}\big(1 &- \kappa_2 f^2(\mu_B) 
+ \kappa_4 f^4(\mu_B)\nonumber \\ 
&-\kappa_6f^6(\mu_B)+\cdots\big)  \,,
\end{align}
where $T_{pc,0}=156.5(1.5)$~MeV is the crossover temperature at $\mu_B=0$ and $f(\mu_B)=\mu_B/T$. 
In recent lattice QCD and other calculations, continuum extrapolated results for the curvature coefficients $\kappa_2$ and $\kappa_4$ were presented and remarkably agreement was found at least for $\mathcal{O}(\mu_B^2)$:  $\kappa_2=0.15(1)$ (for the case $\mu_S=0$) \cite{HotQCD:2018pds,Borsanyi:2020fev,Bellwied:2015rza,Bonati:2018nut,Fu:2019hdw,Gao:2020fbl, Ali:2024nrz}.  
Note that frequently the crossover line is also parametrized with $f(\mu_B)=\mu_B/T_{pc,0}$ and coefficients $\bar\kappa_2$, $\bar\kappa_4$.
We can map one parametrization to the other by setting $\bar\kappa_2=\kappa_2$ and $\bar\kappa_4=\kappa_4-2\kappa_2\bar\kappa_2$, with differences remaining at $\mathcal{O}(\mu_B^6)$. 
Our CEP location for $N_\tau=6$ is roughly in agreement with the $\mathcal{O}(\mu_B^2)$ parametrization in $f(\mu_B)=\mu_B/T$. 
This parametrization is shown in \figref{fig:CEP_histo} as dashed line. 
However, in the region where the CEP is located the contribution of the $\mathcal{O}(\mu_B^4)$ coefficient would already be significant, i.e. the two $\mathcal{O}(\mu_B^2)$ parametrizations with $f(\mu_B)=\mu_B/T$ and $f(\mu_B)=\mu_B/T_{pc,0}$ differ substantially for $\mu_B>400$ MeV or $T<140$ MeV. 
Plugging $\muCEP$ or $\TCEP$ into the latter, would increase $\TCEP$ or $\muCEP$ respectively. 
We note that current cutoff effects that we observe between $N_\tau=6$ and $N_\tau=8$ calculations would lead to a continuum result of $(\muCEP,\TCEP)$ that is consistent with $\bar\kappa_4\approx0$, which is favored by lattice calculations.
Current estimates give $\bar\kappa_4\approx 0.001(7)$ (again for $\mu_S=0$) \cite{HotQCD:2018pds}. 
More precisely the value of $\muCEP$ would increase to $\muCEP\approx 650$ MeV. 
The reason why we are not seriously attempting to perform the continuum extrapolation here is that the $N_\tau=6$ and $N_\tau=8$ calculations differ by their methodology and might suffer from different systematic effects. 

Our estimate for the CEP location (at least when taking cutoff effects into account) compares favorably with a number
of other results and constraints. 
To begin with, it ought to lie outside the estimated convergence region of the Taylor series for $\log\ZQCD$, i.e. it is expected\footnote{Strictly speaking, this estimate of the convergence radius depends on e.g. the temperature. Some of these earlier estimates come from coarse lattices. 
Still, they indicate convergence in the same general regime.} that 
$\mu_B/T\gtrsim 2$~\cite{Giordano:2019slo,Giordano:2019gev,Mukherjee:2019eou,Bollweg:2022rps}.
Moreover, if the CEP exists, it is expected to occur at a lower
temperature than the chiral transition temperature~\cite{Karsch:2019mbv},
which is known to be $132^{+3}_{-6}$ MeV~\cite{HotQCD:2019xnw}. Our estimate
conforms to both of these expectations. Additionally
it is in rough agreement with recent predictions from
Dyson-Schwinger equations~\cite{Gao:2020fbl,Gunkel:2021oya},
the functional renormalization group~\cite{Fu:2019hdw},
and black-hole engineering~\cite{Hippert:2023bel}.
Similar to our approach, the conformal Pad\'e applied to the
same HotQCD data yields a compatible result as well~\cite{Basar:2023nkp}.

An important limitation of the estimates presented here\footnote{And therefore
also for Ref.~\cite{Basar:2023nkp}.} is that they are not fully extrapolated to the continuum limit. 
For the multi-point Pad\'e, we are limited to $N_\tau=6$,
and the HotQCD data utilize $\chi_6^B$ and $\chi_8^B$ on $N_\tau=8$
lattices. 
At this stage, it appears clear that both data sets are sensitive
to a non-trivial singularity structure in $\mu_B$ that is consistent with Lee-Yang edge scaling, but in principle our estimates of the location of $\muCEP$ are distorted by cutoff effects.

\textit{Summary.}---Here we present a new strategy for the critical point search at $\mu_B>0$ by means of first principle lattice QCD calculations. Based on rational function approximations of the cumulants of the baryon-number fluctuations at imaginary chemical potential we determine singularities in the complex $\mu_B$ plane. We extrapolate these singularities using a scaling ansatz motivated by the temperature scaling of the Lee-Yang edge singularity. The rational function approximations are obtained by a multi-point Pad\'e analysis on a sliding interval embedded in $\mu_B/T\in[-i\pi,i\pi]$.
For the $N_\tau=6$ results we find $\TCEP=105^{+8}_{-18}$~ MeV and $\muCEP = 422^{+80}_{-35}$~ MeV, based on $\mathcal{O}(10^5)$ different fits. However we expect that cutoff effects will alter $\muCEP$ towards larger values of around $\sim650$ MeV. This estimate is roughly consistent with other estimates in the literature and also with the current determination of the crossover line. 

\textit{Acknowledgement.}---This work was supported by the Deutsche Forschungsgemeinschaft (DFG, German Research Foundation) Proj. No. 315477589-TRR 211, by the PUNCH4NFDI consortium supported by the Deutsche Forschungsgemeinschaft (DFG, German Research Foundation) with project number 460248186 (PUNCH4NFDI) and by INFN (Istituto Nazionale di Fisica Nucleare) under research project {\em i.s. QCDLAT}. In its early phase this work also  received funding from the European Union’s Horizon 2020 research and innovation programme under the Marie Skłodowska-Curie grant agreement No. 813942 ({\em EuroPLEx}).
Numerical calculations have been made possible
through EuroHPC JU and PRACE grants at CINECA, Italy on Leonardo and Marconi100, and through the Gauss Centre for Supercomputing e.V. (www.gauss-centre.eu) on the GCS Supercomputer JUWELS \cite{JUWELS} at Jülich Supercomputing Centre (JSC). 
Additional calculations have been performed on Leonardo under the INFN-CINECA agreement for HPC and on the GPU clusters at Bielefeld University, Germany. We also acknowledge support of the Bielefeld NPC.NRW team. DAC was supported by the National Science Foundation under Grant PHY20-13064. KZ acknowledges support by the project ``Non-perturbative aspects of fundamental interactions, in the Standard Model and beyond” funded by MUR, Progetti di Ricerca di Rilevante Interesse Nazionale (PRIN), Bando 2022, Grant 2022TJFCYB (CUP I53D23001440006).

\appendix

\bibliography{bibliography}
\clearpage
\section*{supplemental material}
\begin{figure*}[h]
    \centering
    \includegraphics[width=0.45\linewidth]{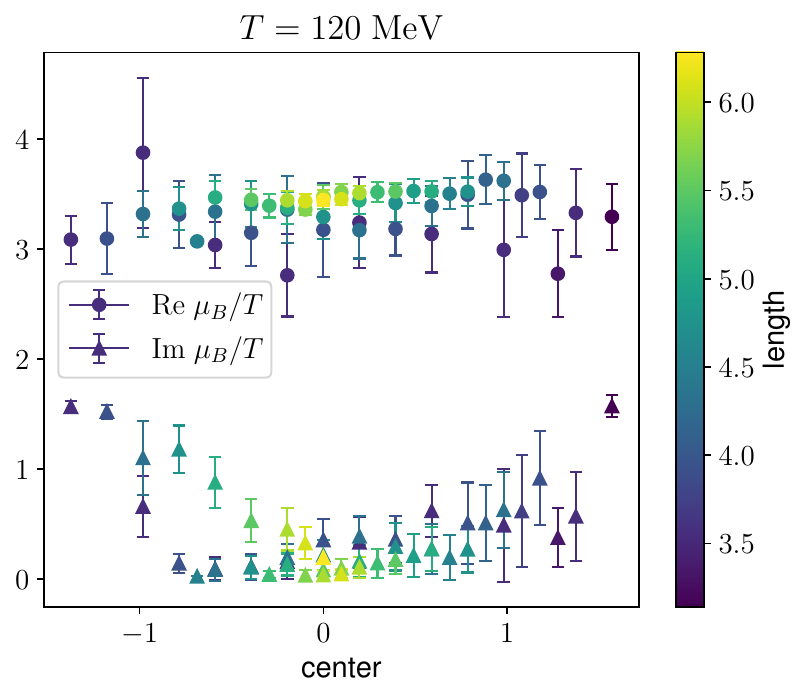}\hfill
    \includegraphics[width=0.45\linewidth]{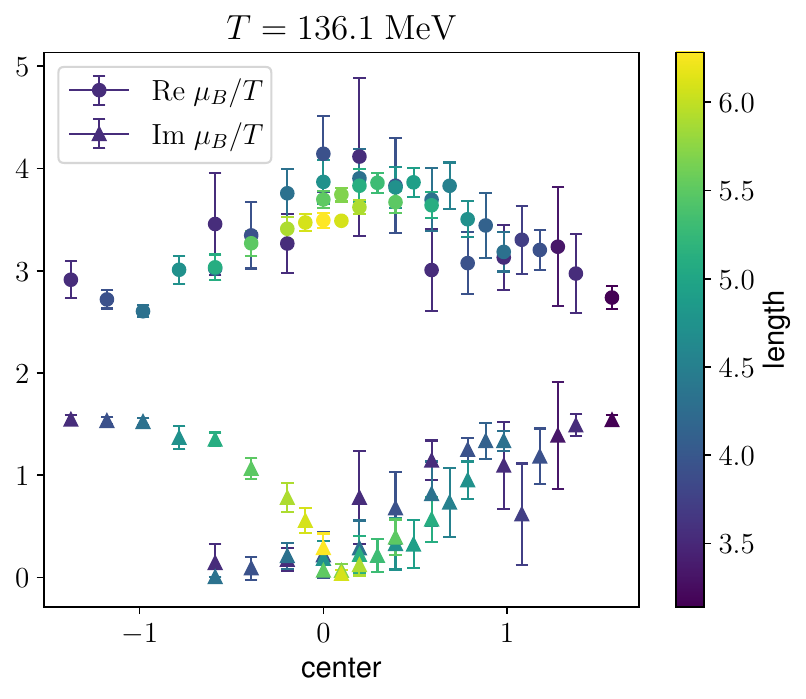}
    \caption{Fit range dependence for the rational approximation $R_3^3(\mu_B/T)$ for $T=120$~MeV (left) and $T=136$~MeV (right). Shown are the real (circles) and imaginary (triangles) parts of the obtained poles vs. the center of the fit interval. The length of the fit interval is indicated by the color and varies from $\pi$ to $2\pi$.}
    \label{fig:interval_dependence}
\end{figure*}

\textit{The sliding window analysis of the rational function approximation.}--As we had already observed in Ref.~\cite{Dimopoulos:2021vrk}, there is an interval dependence for the poles located by multi-point Pad\'e analysis. When we change the interval we become more sensitive to some singularities and less to others. In particular we observe that for symmetric intervals centered at $\mu_B/T=i\pi$, thermal singularities that belong to the Roberge-Weiss transition are favored. For that reason we perform a very general sliding window analysis where we vary the interval length as well as the center of the interval used in the rational approximation. The length is varied between $\pi$ and $2\pi$, whereas the center of the interval is located in [$-i\pi/2$, $i\pi/2$]. Note that only the points between $0$ and $i\pi$ are calculated. In other regions the data is duplicated through reflection symmetry and periodicity. The procedure is visualized in \figref{fig:sliding_window}, for the example of the $T=120$ MeV data and can be summarized by the following algorithm:
\begin{figure}
    \centering
    \includegraphics[width=\linewidth]{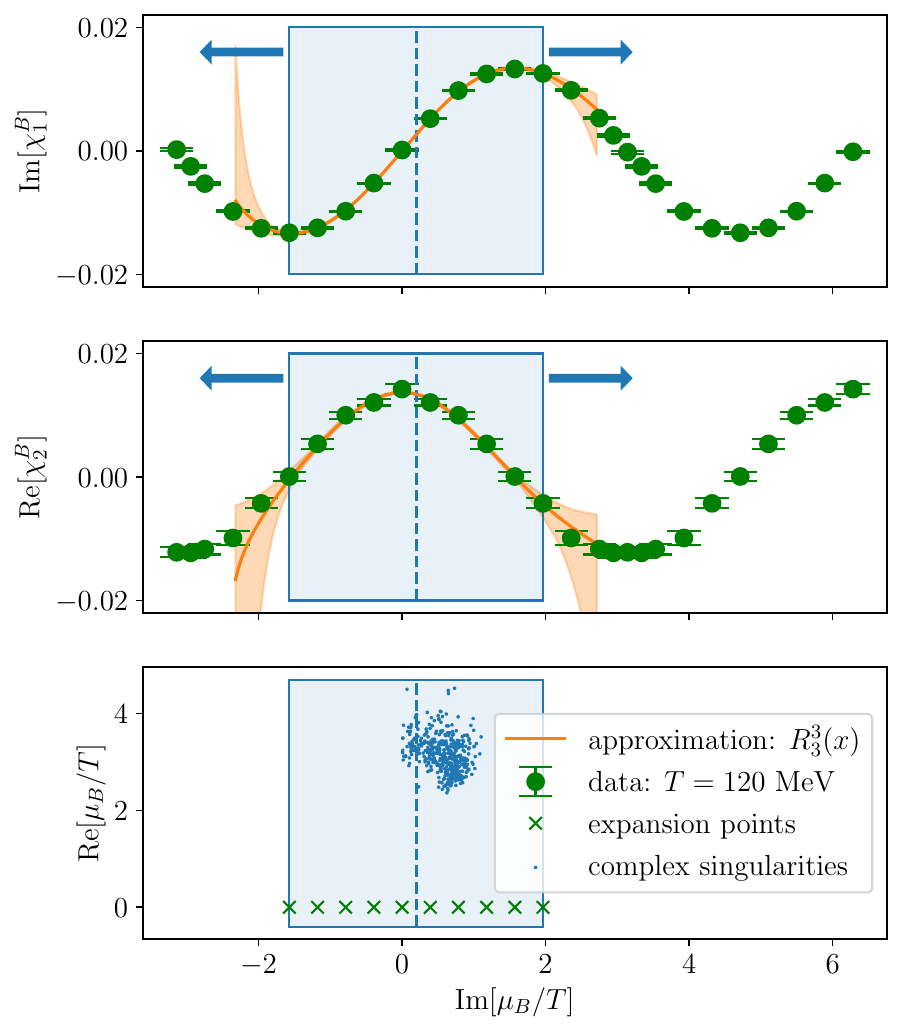}
    \caption{A visualization of the sliding window analysis for $T=120$~MeV. We show the data for $\chi_1^B$ (upper panel) and $\chi_2^B$ (middle panel) together with the rational approximation $R_3^3(\mu_B/T)$ obtained from the fit interval that is indicated by the blue box. The error band of the fit is obtained by Gaussian bootstrapping. The complex singularities obtained for each of the bootstrap samples are shown on the lower panel.}
    \label{fig:sliding_window}
\end{figure}
\begin{algorithm}[H]
\SetKwData{Ii}{$I_i$}
\SetKwData{Dj}{$D_{j}$}
\SetKwData{Rj}{$R_{j}$}
\SetKwData{Sj}{$S_{j}$}
\SetKwData{Mi}{$M_{i}$}
\SetKwData{Ei}{$E_{i}$}
\SetKwFunction{SelectInterval}{SelectInterval}
\SetKwFunction{DrawBootstrapSample}{DrawBootstapSample}
\SetKwFunction{RationalApproximation}{RationalApproximation}
\SetKwFunction{CalcSingularities}{CalcSingularities}
\SetKwFunction{CalcMean}{CalcMean}
\SetKwFunction{CalcError}{CalcError}
\SetKwFunction{ClearData}{ClearData}
\Begin{
  \ClearData{M,E}\;
  \For{$i \in N_{\rm interval}$}{
    \Ii $\leftarrow$ \SelectInterval{i}\;
    \ClearData{D,R,S}\;
    \For{$j \in N_{\rm samples}$}{
       \Dj $\leftarrow$ \DrawBootstrapSample{\Ii}\;
       \Rj $\leftarrow$ \RationalApproximation{\Dj}\;
       \Sj $\leftarrow$ \CalcSingularities{\Rj}\;
    }
    \Mi $\leftarrow$ \CalcMean{S}\;
    \Ei $\leftarrow$ \CalcError{S}\;
  }
}
\caption{sliding window analysis}\label{alg:swa}
\end{algorithm}
The subroutine \texttt{SelectInterval}$(i)$ is chosen to be deterministic. 
We number all possible combinations of interval length and center in the above mentioned ranges that are possible with our data points. We find $N_{\rm interval}=55$. 
The subroutine \texttt{DrawBootstrapSample}$(I_i)$ on the other hand assumes independent and normal distributed errors on the data points that have support in the interval $I_i$.
The subroutine \texttt{RationalApproximation}$(D_j)$ performs a combined fit to the $\chi_1^B$ and $\chi_2^B$ data from the sample $D_j$. 
We use the rational function $R_3^3(x)$ as given in \equatref{eq:ratapprox}. We checked that higher order rational approximations give very similar results, however, the initial set of parameters for these fits have to be chosen more carefully to ensure convergence, and cancellations of poles in the numerator and denominator are more likely to occur. We keep only approximations with $\chidof<2.5$. In \texttt{CalcSingularities }$(R_j)$, we determine the roots of the denominator of $R_3^3$ and check if they are canceled by poles in the numerator. For such cancellations we allow for a tolerance of $\Delta=0.03$.
Finally we calculate the mean and error for the singularities for which we chose the median and 1-$\sigma$ percentile of the real and imaginary part. For the ellipses shown in \figref{fig:confelipse} we also calculate the Pearson correlation coefficient. We repeat the sliding window analysis for each of the temperatures. The results for the real and imaginary parts of the singularities at the example of $T=120$ MeV  and $T=136$~MeV are shown in \figref{fig:interval_dependence}.
\begin{figure*}[tbh]
    \centering
    \includegraphics[height=0.35\linewidth]{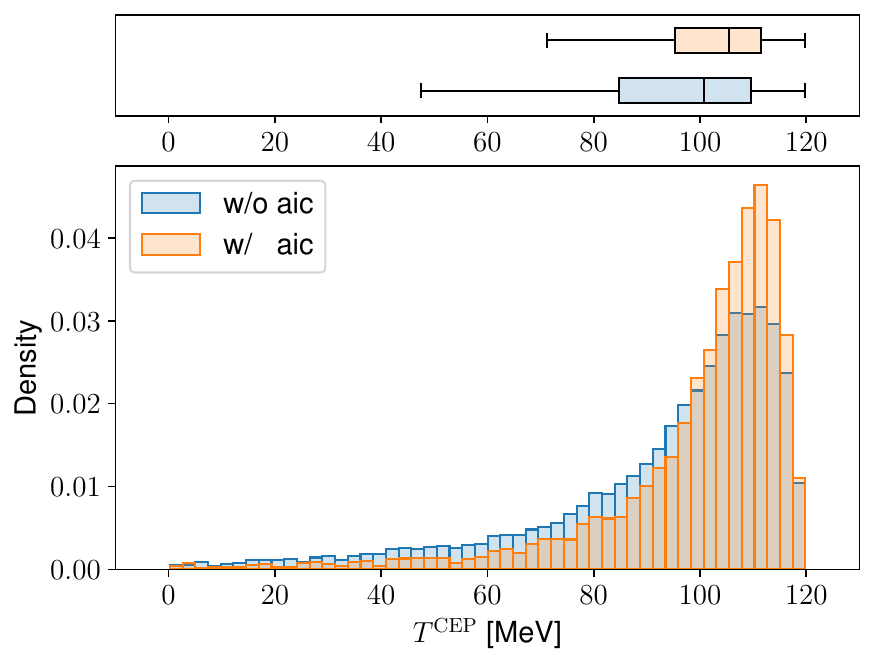}\hfill
    \includegraphics[height=0.35\linewidth]{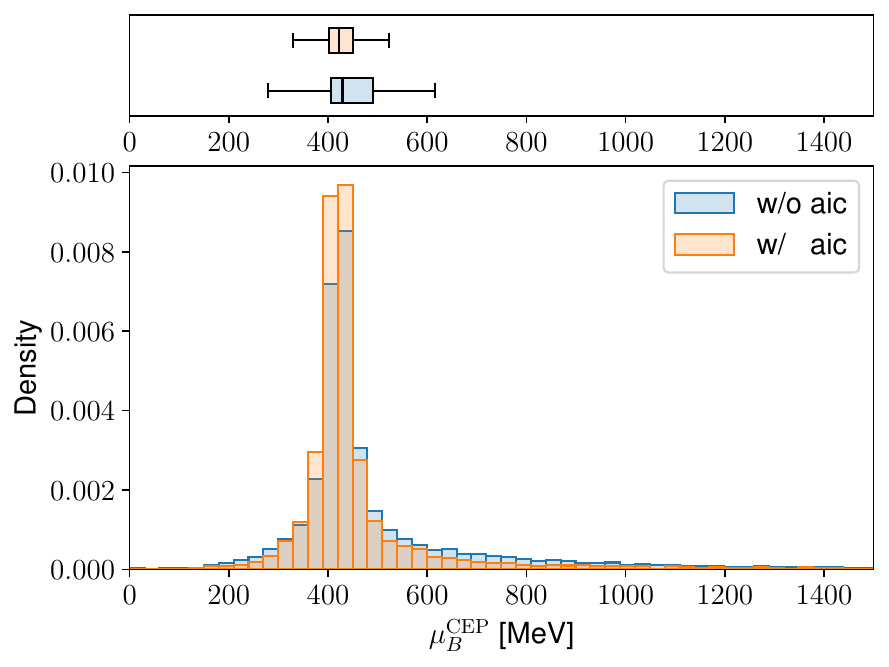}
    \caption{Probability density of $\TCEP$ (left) and $\muCEP$ (right) with standard boxplots on top of the data, based on $\mathcal{O}(10^5)$ individual fits of \eqref{eq:CEPfit} to the positions of the LYE.}
    \label{fig:1d_hist}
\end{figure*}

\textit{Statistical analysis of fit results.}--
Next we perform scaling fits to equation \eqref{eq:CEPfit}. We chose one of the 55 intervals for each of the 5 different temperatures. This gives us $55^5$ different data sets. In practice we chose $\mathcal{O}(10^5)$ random samples from possible interval combinations. The results for $\TCEP$ and $\muCEP$ are summarized in \figref{fig:CEP_histo} as a 2d-histogram. We also calculate the relative likelihood of each fit in accordance with the Akaike information criterion. In particular, we calculate the weights
\begin{equation}
    w_{{\rm AIC},j}
    = \frac{ \exp\{ ( {\rm AIC}_{\rm min} - {\rm AIC}_{j} )/2\} }{ \sum_j \exp\{({\rm AIC}_{\rm min}-{\rm AIC}_j)/2\}}\;,
\end{equation}
where index $j$ labels the fits and ${\rm AIC}=2\log\mathcal{L}=\chi^2$. We observe that models with small $\TCEP$ or large $\muCEP$ are further suppressed when we weight the results with $w_{{\rm AIC},i}$. A one dimensional histogram of the $\TCEP$ and $\muCEP$ results, respectively, is shown in \figref{fig:1d_hist}. 
On top of the histograms we show standard box-and-whiskers diagrams for the data, where the vertical line inside the boxes indicates the median. Note that the size of the box indicates here the interquartile range (IQR=$Q_3$-$Q_1$) and the whiskers have the length of 1.5$\cdot$IQR, except for $T>\TCEP$, where we indicate the maximal values that occur. 
\end{document}